\documentclass[aps,prl,twocolumn,showpacs,groupedaddress,affilletter,nobibnotes,nofootinbib]{revtex4}

\usepackage{graphicx}
\usepackage{epsfig}

\newcommand{\columnwidthstrach}{2.5in}

\begin{document}


\title{Controlled Fabrication of Nanogaps in Ambient Environment for Molecular Electronics}


\author{D.~R. Strachan,$\mathrm{^{a),b),}}$\footnote{Electronic Address:
drstrach@sas.upenn.edu} D. E. Smith,$\mathrm{^{a)}}$ D. E.
Johnston,$\mathrm{^{a)}}$ T.-H. Park,$\mathrm{^{c)}}$ Michael J.
Therien,$\mathrm{^{c)}}$ D. A. Bonnell,$\mathrm{^{b)}}$ and A. T.
Johnson$\mathrm{^{a),}}$\footnote{Electronic Address:
cjohnson@physics.upenn.edu}}
\affiliation{$\mathrm{^{a)}}$Department of Physics and Astronomy,
University of Pennsylvania, Philadelphia, PA 19104 \\
$\mathrm{^{b)}}$Department of Materials Science and Engineering,
University of Pennsylvania, Philadelphia, PA 19104  \\
$\mathrm{^{c)}}$Department of Chemistry, University of
Pennsylvania, Philadelphia, PA 19104}



\date{\today}

\begin{abstract}
We have developed a controlled and highly reproducible method of
making nanometer-spaced electrodes using electromigration in
ambient lab conditions.  This advance will make feasible single
molecule measurements of macromolecules with tertiary and
quaternary structures that do not survive the liquid-helium
temperatures at which electromigration is typically performed. A
second advance is that it yields gaps of desired tunnelling
resistance, as opposed to the random formation at liquid-helium
temperatures.  Nanogap formation occurs through three regimes:
First it evolves through a bulk-neck regime where electromigration
is triggered at constant temperature, then to a few-atom regime
characterized by conductance quantum plateaus and jumps, and
finally to a tunnelling regime across the nanogap once the
conductance falls below the conductance quantum.
\end{abstract}


\maketitle

Electromigration has recently been successfully employed to make
nanometer-spaced electrodes for single molecule
devices~\cite{Liang02_1nature,Park02_1nature,Park00_1nature,Yu04_1nano}.
The typical procedure entails an abrupt break at liquid-helium
temperatures that yields a nanogap with a random tunnelling
resistance~\cite{Yu04_1nano,Park99_1apl,Bolton04_1apl,Selzer04_1nanotech,Lambert03_1nanotech}.
However, this procedure makes gaps at room temperature which are
typically too large for molecular
measurements~\cite{Bolton04_1apl}.  This hinders the application
of the typical electromigration procedure to molecules which do
not survive a sub-freezing environment, such as macromolecules
that feature modest thermodynamic stability of their respective
tertiary and quaternary structures.

We have developed an electromigration procedure that is completely
performed in ambient laboratory conditions and yields a
controllable nanogap resistance to within a factor of about three
of the target value in the 0.5 M$\Omega$ to 1 T$\Omega$ range. The
electromigration procedure evolves through three regimes. At large
conductance ($G$), local heating increases Au mobility and
triggers electromigration in the metallic neck at a critical
temperature. When the neck narrows to the few-atom regime it shows
jumps and plateaus near multiples of the conductance quantum
($G_o={{2e^2}/{h}}$) and a sharp decrease in the critical
temperature. A tunnelling regime is entered once $G$ falls below
$G_o$ accompanied by formation of a nanogap.

\begin{figure}
\epsfig{file=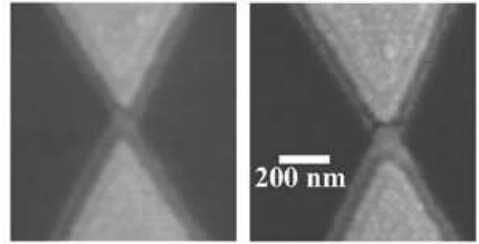,clip=,width=\columnwidthstrach}
\caption{\label{gap_fig} (a) Field-emission SEM micrograph of
electrodes before electromigration. (b) Nanogap after
electromigration.}
\end{figure}

We first fabricate two overlapping Au leads (each 8-30 nm thick)
using electron-beam lithography and double-angle evaporation of Au
(Fig.\ \ref{gap_fig}a).  An initial 3 nm thick Cr layer (deposited
normal to surface) helps the contact pads adhere to the
$\mathrm{SiO_{2}}$ substrate while a final 40 nm thick layer of Au
on the contacts reduces the resistance to between 100-200 $\Omega$
at room temperature.

\begin{figure}
\epsfig{file=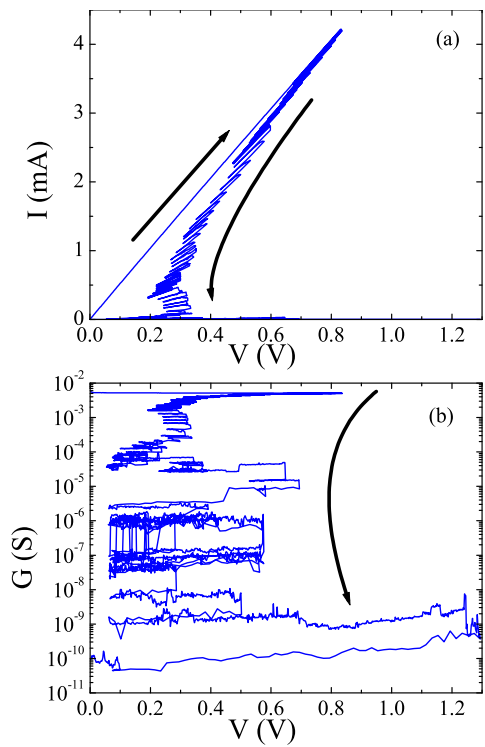,clip=,width=\columnwidthstrach}
\caption{(a) $I-V$ data from a succession of ramps across a
junction. Arrows indicate how $I-V$ evolves as the nanogap forms.
The ramp up shows the initial ohmic behavior of leads and the
downward pattern is the electromigration. (b) Same data with $G$
plotted on $\log$ scale.} \label{IV_G_break}
\end{figure}

At room temperature and atmospheric pressure, we perform
controlled electromigration with a succession of voltage ($V$)
ramps while monitoring the current ($I$) and conductance of the
leads (Fig.\ \ref{IV_G_break}).  We make an initial measurement of
$G$ and compare this to later measurements as $V$ is ramped up 4
mV/sec. When $G$ decreases by a certain threshold percentage
($Th$) we ramp $V$ down about 100 mV at 40 mV/sec to arrest the
breaking, and then we repeat the ramping procedure to slowly form
the nanogap. (We typically start with $Th=$1\% and gradually
increase this up to 150\% as the nanogap is formed.)

The gap conductance can be made to fall in a controlled manner to
within a factor of about three of the desired value.  Figure\
\ref{IV_G_break}b shows the break to about 10 G$\Omega$. Figure\
\ref{gap_fig}b is a micrograph of a typical junction after
performing the procedure.  This demonstrates that a gap has
clearly formed, although the location of closest approach between
the electrodes is not resolved.

The reproducibility and control of the procedure permit it to be
completely automated.  It currently takes us between one and two
hours to form a nanogap with a final resistance of about 10
G$\Omega$. Although $Th$ can be slightly increased to form
nanogaps more quickly, this seriously degrades the reproducibility
of the procedure and often results in catastrophic breaks with
gaps greater than 100 nm.

\begin{figure}
\epsfig{file=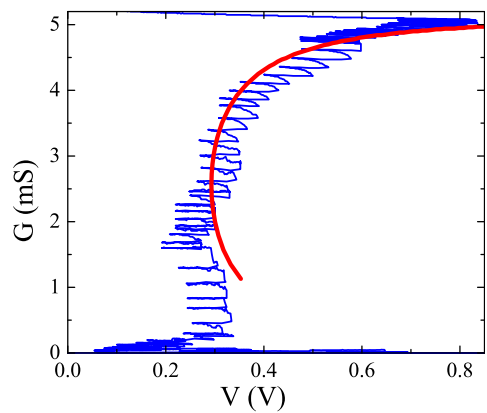,clip=,width=\columnwidthstrach}
\caption{Fit of Joule-heating model (Eq.\ \ref{break_eq2}) to the
electromigration of Au leads.} \label{break_evolution}
\end{figure}

The slow gap formation permits an investigation of
electromigration~\cite{christou94_1,Ho89_1rep_prog_phys}. At early
stages, thermal heating accounts for the triggering of
electromigration in the metallic neck.  As the cross-section of
the metallic neck decreases, its resistance ($R_{n}$) increases.
Thus, $V$ across the neck and leads is $V=I \left( R_L + R_n
\right)$, where $R_L$ is the resistance of the leads. Assigning
${P_n}^*=R_n{I^*}^2$ as the critical power, the critical voltage
is
\begin{equation}
V^*=\sqrt{{{{P_n}^*}\over{G(1-G R_L)}}}, \label{break_eq2}
\end{equation}
with $G$ the conductance.  $R_L$ can be measured independently
before starting to break (as $R_L >> R_n$ at this stage) so the
only fitting parameter is the critical power.  In Fig.\
\ref{break_evolution}, we fit Eq.\ \ref{break_eq2} to the data
using ${P_n}^*=110 \mu$W. We have likewise fit data from 20 other
gaps and find good agreement to the model as long as the gap
conductance is larger than 1 mS (with ${P_n}^* \approx 200 \mu$W
$\pm 100 \mu$W).  This indicates that Joule heating increases the
mobility of the Au atoms in this bulk-neck regime which triggers
the smoothly evolving electromigration.

\begin{figure}
\epsfig{file=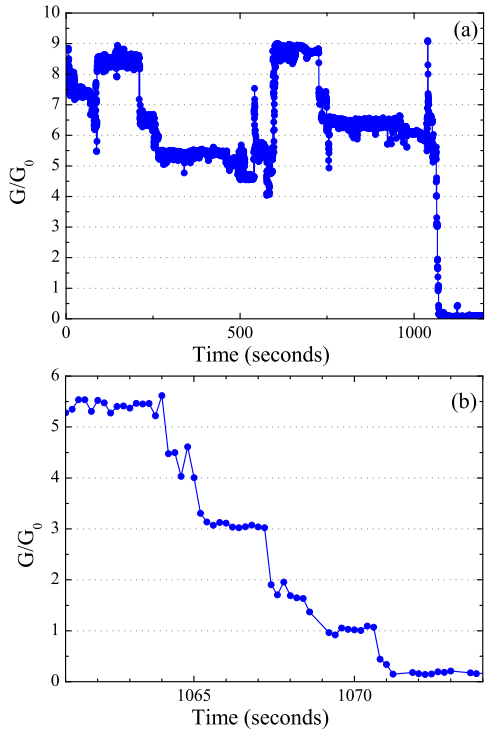,clip=,width=3.1in} \caption{(a) Jumps and
plateaus in $G$ of a metallic neck formed to 1.2 k$\Omega$ with
the controlled electromigration procedure.  The measurements were
performed at 89 mV.  Time equal to zero corresponds to the point
when the voltage reached 89 mV.  (b) Expanded view of data at
location where $G$ falls below $G_o$. } \label{steps2}
\end{figure}

After $G$ decreases below 1 mS, electromigration occurs before
reaching ${P_n}^*$ and the smooth evolution changes to one
characterized by discrete jumps to plateaus both up and downwards
by approximately $G_o$.  In contrast to methods using continuous
mechanical strains (\emph{e.g.}, Refs.\
\onlinecite{Muller92_1prl,Muller96_1prb,Krans95_1nature,Olesen94_1prl,Pascual95_1science}),
our procedure naturally leads to plateaus when ramping down which
makes it difficult to discern those plateaus due to an
atomic-sized neck. To circumvent this issue, Fig.\ \ref{steps2}a
shows $G$ as a function of time for a junction immediately after
controlled electromigration to 1.2 k$\Omega$, where the
measurements were made with the voltage fixed at 89 mV. In such
samples, we find that the conductance undergoes jumps to plateaus
spaced roughly $G_o$ apart.  As with the mechanical break junction
technique~\cite{Muller96_1prb}, the plateaus when $G \geq 5 \times
G_o$ often do not correspond to integral multiples of $G_o$.

As $G$ falls towards $G_o$, plateaus near multiples of $G_o$
appear. This is clearly demonstrated in Fig.\ \ref{steps2}b, which
shows an expanded view of the data at the location where $G$
approaches $G_o$.  These plateaus are reminiscent of the
quantization found with the mechanical break junction
technique~\cite{Muller96_1prb}.  Similar to the mechanical break
junction technique, we sometimes find plateaus for $G < 5 \times
G_o$ which do not correspond to integral multiples of $G_o$,
though our technique generally shows reasonable agreement with
quantized conductance. Quantization implies that the metallic neck
is only a few atoms wide in this regime and that a nanogap is
formed once $G$ is below $1 \times G_o$. We also generally find
that passing below $G_o$ is the least controllable stage of the
procedure, with the conductance sometimes abruptly falling by as
much as a factor of 100 from a value near $G_o$. This is further
evidence that transport shifts over to tunnelling in the final
nanogap regime.

It is not surprising that electromigration in the few-atom regime
does not occur at ${P_n}^*$, as details of atomic structure and
bonding are important in this
case~\cite{Rubio96_1prl,Burki03_1prl}.  As $G$ approaches $G_o$, a
deviation from Eq.\ \ref{break_eq2} towards lower $V$ is clearly
seen in the data of Fig.\ \ref{break_evolution} and for those
obtained from all other gaps we have made.  It is likely that this
failure is due to the simplicity of our model, which assumes that
the dissipation occurs only in the neck and that the heat is
transported away through a constant thermal conductance.

In conclusion, we have developed an electromigration procedure
that can be completely performed in ambient conditions and yields
a controllable nanogap resistance.  This advance could
considerably aid formation of electrical contacts to molecules
which cannot survive sub-freezing conditions.  The procedure
evolves through a bulk-neck regime where electromigration is
triggered at constant temperature, then to a few-atom regime
characterized by quantum of conductance plateaus and jumps, and
finally to a tunnelling regime once the gap is formed.

This work was supported through NSF-NIRT grant 0304531 and
MRSEC-NSF grant DMR-00-79909.  DJ acknowledges financial support
from NSF IGERT program (grant DGE-0221664) and SENS.

\bibliography{nano_refs}



\end{document}